\documentclass{article}
\usepackage{emulateapj,pstricks,apjfonts}

\def\1e{\mbox{1E0657--56}}
\def\einstein   {{\em Einstein}\/}
\def\asca       {{\em ASCA}\/}
\def\rosat      {{\em ROSAT}\/}
\def\chandra    {{\em Chandra}\/}

\def\as         {$^{\prime\prime}$}
\def\ergs       {~erg$\;$s$^{-1}$}
\def\kms        {~km$\;$s$^{-1}$}
\def\kmsmpc     {~km$\;$s$^{-1}\,$Mpc$^{-1}$}
\def\cmsq       {~cm$^{-2}$}
\def\gax        {\gtrsim}
\def\deg        {$^{\circ}$}
\def\bi{\bfseries\itshape}

\begin{document}

\submitted{ApJ Letters in press; astro-ph/0110468 v2}

\lefthead{MERGER SHOCK IN 1E0657--56}
\righthead{MARKEVITCH ET AL.}

\title{A TEXTBOOK EXAMPLE OF A BOW SHOCK IN THE MERGING GALAXY CLUSTER
1E0657--56}

\author{M.~Markevitch, A. H. Gonzalez, L.~David, A.~Vikhlinin, S.~Murray,
W.~Forman, C.~Jones, W.~Tucker$^1$}

\affil{Harvard-Smithsonian Center for Astrophysics, 60 Garden St.,
Cambridge, MA 02138; maxim@head-cfa.harvard.edu}

\footnotetext[1]{Also at UCSD.}
\setcounter{footnote}{1}

\begin{abstract}

The \chandra\ image of the merging, hot galaxy cluster \1e\ reveals a bow
shock propagating in front of a bullet-like gas cloud just exiting the
disrupted cluster core. This is the first clear example of a shock front in
a cluster.  From the jumps in the gas density and temperature at the shock,
the Mach number of the bullet-like cloud is 2--3. This corresponds to a
velocity of 3000--4000 \kms\ relative to the main cluster, which means that
the cloud traversed the core just 0.1--0.2 Gyr ago. The 6--7 keV
``bullet'' appears to be a remnant of a dense cooling flow region once
located at the center of a merging subcluster whose outer gas has been
stripped by ram pressure. The bullet's shape indicates that it is near the
final stage of being destroyed by ram pressure and gas dynamic
instabilities, as the subcluster galaxies move well ahead of the cool
gas. The unique simplicity of the shock front and bullet geometry in \1e\
may allow a number of interesting future measurements. The cluster's average
temperature is $14-15$ keV but shows large spatial variations. The hottest
gas ($T>20$ keV) lies in the region of the radio halo enhancement and
extensive merging activity involving subclusters other than the bullet.

\end{abstract}


\section{INTRODUCTION}

Galaxy clusters form via mergers of smaller subunits. Such mergers dissipate
a large fraction of the subclusters' vast kinetic energy through gas dynamic
shocks, heating the intracluster gas and probably accelerating high energy
particles (e.g., Sarazin 2001 and references therein).
Shocks contain information on the velocity and geometry of the merger. They
also provide a unique laboratory for studying the intracluster plasma,
including such processes as thermal conduction and electron-ion
equilibration (e.g., Shafranov 1957; Takizawa 1999). Some exploratory uses
of X-ray data on cluster shocks were described by Markevitch, Sarazin, \&
Vikhlinin (1999). While many merging clusters exhibit recently heated gas
(e.g., Henry \& Briel 1995; Markevitch et al.\ 1999, Furusho et al.\ 2001
and references in those works; Neumann et al.\ 2001), so far only two
candidate merger shock fronts were observed. One is a mild X-ray brightness
edge, apparently a shock with a Mach number near 1, preceding the prominent
``cold front'' in A3667 (Vikhlinin, Markevitch \& Murray 2001).  Another is
a hot region in front of the A665 core (Markevitch \& Vikhlinin 2001,
hereafter MV) which shows no clear density jump perhaps because of an
unfavorable viewing geometry.

The \chandra\ observation of \1e\ presents the first clear example of a
cluster bow shock. This $z=0.296$ cluster was discovered by Tucker,
Tananbaum, \& Remillard (1995) as an \einstein\ IPC extended source.  From
\asca\ data, Tucker et al.\ (1998, hereafter T98) derived a temperature
around 17 keV, making this system one of the hottest known (see also Yaqoob
1999; Liang et al.\ 2000, hereafter LHBA). \rosat\ data showed that \1e\ is
a merger (T98). It also hosts the most luminous synchrotron radio halo
(LHBA).

Below we present results from the \chandra\ observation of \1e\ performed in
October 2000. We use $H_0=100\,h$\kmsmpc\ and $\Omega_0=0.3$, $\Lambda=0$;
$1'=0.172\,h^{-1}$ Mpc at the cluster redshift. Confidence intervals are
90\% for one-parameter, unless specified otherwise.

\begin{figure*}[t]
\pspicture(0,14.2)(18.5,24)

\rput[tl]{0}(-0.1,24.0){\epsfxsize=9.4cm \epsfclipon
\epsffile{1e_gray.ps_dist}}

\rput[tl]{0}(9.2,24.0){\epsfxsize=9.4cm \epsfclipon
\epsffile{1e_sm2cont.ps_dist}}

\rput[bl]{0}(8.2,23.3){\large\bi a}
\rput[bl]{0}(17.6,23.3){\large\bi b}

\rput[tl]{0}(-0.1,15.3){
\begin{minipage}{18.5cm}
\small\parindent=3.5mm
{\sc Fig.}~1.---({\em a}) ACIS 0.5--5 keV image. Pixels are 4\as; linear
scale is for $h=0.5$. ({\em b}) Grayscale shows optical $R$-band image from
ESO NTT (courtesy of E. Falco and M. Ramella). Contours, spaced by a factor
of 2, show the smoothed ACIS image. The western part of the outer contour
is approximately at the shock position. A number of well-matched point
sources shows that the coordinates are accurate.
\par
\end{minipage}
}
\endpspicture
\end{figure*}

\section{DATA ANALYSIS}
\label{sec:analysis}

\1e\ was observed by ACIS-I at the focal plane temperature of $-120^\circ$C
for a useful exposure of 24.3 ks. To derive the gas temperature for a given
region of the cluster image, the telescope and detector response were
modeled as described in MV. The ACIS background rate did not vary during the
exposure, but was higher than expected by a factor of about 1.3, most likely
due to anomalous ``space weather''. This required special background
modeling. To do this, we extracted a spectrum from the ACIS-I region outside
an $r=8.6'$ ($1.5\,h^{-1}$ Mpc) circle centered on the cluster, which should
be free of cluster emission. Point sources were excluded. The observed
excess over the nominal background%
\footnote{A combination of blank field observations normalized by the
exposure time, see http://asc.harvard.edu/cal, click ``ACIS'', then ``ACIS
Background''}
was well-modeled in the $0.7-10$ keV band by the sum of two power laws
$E^\alpha$ with photon indices $\alpha=-0.6$ and $+3.0$ (dominant below and
above $E\sim 5$ keV, respectively) originating inside the detector, i.e.,
without applying mirror effective area and CCD efficiency to the model. Such
a background anomaly in the ACIS-I chips is rare and not yet understood. We
assumed that this component is distributed uniformly over the detector, and
added it (normalized by solid angle) to the nominal background
spectra. High-energy residuals in the overall cluster fit indicated that the
normalization of the corrected background required an additional 10\%
increase (perhaps indicating some spatial nonuniformity of the excess),
which we applied to the spectra from all cluster regions.

To make a $0.5-5$ keV image for the gas density analysis, we compared the
observed background rate far from the cluster to the nominal model
background (without the above additional component) and derived a correction
factor of 1.35 for this wide band, which was applied to the model background
image. This is of course consistent with the above spectral correction. A
10\% background uncertainty was included in deriving the confidence
intervals for all quantities. This approximate background modeling is
adequate for our present qualitative study.

To minimize the effects of calibration uncertainties, the spectra were
extracted in the $0.9-9.5$ keV energy band, excluding the $1.8-2.2$ keV
interval around the mirror Ir edge.

\section{RESULTS}
\label{sec:res}

The ACIS image of the cluster is shown in Fig.\ 1{\em a}. It shows a
``bullet'' apparently just exiting the cluster core and moving westward.
This subcluster was previously seen in the \rosat\ data (e.g., T98), but the
high resolution \chandra\ image makes its nature and direction of motion
clear. The bullet is preceded by an X-ray brightness edge that resembles a
bow shock. To determine whether it is indeed a shock (and not a ``cold
front'', e.g., Markevitch et al.\ 2000; Vikhlinin et al.\ 2001), below we
derive the gas temperatures on both sides of the feature (see
\S\ref{sec:tmap}). Figure 1{\em b}\/ shows X-ray contours overlaid on an
optical image.  A subcluster of galaxies (e.g., Barrena et al.\ 2001) is
seen leading the X-ray bullet, which is apparently swept back from the
galaxies by the ram pressure of the ambient cluster gas.

\subsection{Temperature Map}
\label{sec:tmap}

We first fit an overall cluster spectrum within $r=3'$ ($0.5\,h^{-1}$ Mpc)
as described in \S\ref{sec:analysis} using the Kaastra (1992) plasma
model. We obtain $T=14.8^{+1.7}_{-1.2}$ keV and an abundance of $0.11\pm
0.11$ solar, fixing the absorption at $N_H=4.6\times 10^{20}$\cmsq\ as
derived by LHBA from radio and \rosat\ PSPC data. This temperature agrees
with their \asca\ + \rosat\ fit of $14.5^{+2.0}_{-1.7}$ keV and is
consistent with the $17.4\pm 2.5$ keV \asca\ fit by T98. If $N_H=6.5\times
10^{20}$\cmsq\ from Dickey \& Lockman (1990) is used instead, we obtain
$T=13.6$ keV. When $N_H$ is fit as a free parameter, it is consistent with
both those values; the present low-energy ACIS-I calibration is not reliable
for measuring $N_H$ independently. Below we fix $N_H=4.6\times
10^{20}$\cmsq. Within the $r=0.5\,h^{-1}$ Mpc aperture, the cluster's 0.5--5
keV (rest-frame) luminosity is $8.7\pm 0.5 \times 10^{44}\, h^{-2}$\ergs\
and $L_{\rm bol}=2.3\pm 0.2\times 10^{45}\, h^{-2}$\ergs.

Figure 2 shows a temperature map made by dividing the cluster image into
several regions and fitting their temperatures and abundances as
above. Despite large uncertainties, the map shows that the gas outside the
shock feature (region {\em P}) is cooler, or at least not hotter, than that
inside (region {\em S}), confirming that it is indeed a shock front. The
temperature at the tip of the bullet is low ($\sim 7$ keV) and is likely to
have been the temperature of the subcluster. The hottest region of the
cluster is its southeastern X-ray brightness elongation. The optical image
shows several large galaxies in that area, suggesting that this is the main
merger site. As seen in Fig.\ 3, this also is where the radio halo is
enhanced (LHBA).  The halo also extends to the western shock front. A
spatial correlation between the halo brightness and the local gas
temperature (in addition to the general similarity to the X-ray brightness,
e.g., LHBA; Govoni et al.\ 2001) was noticed by MV in two other merging
clusters and supports the merger shock origin for the relativistic halo
electrons (e.g., Tribble 1993).

\section{DISCUSSION}

\subsection{The shock front}

The temperature map confirms that the western X-ray brightness edge is a
shock front. We can derive its Mach number from either the temperature or
density jump across the front using the Rankine-Hugoniot shock adiabat.
Figure 4{\em a} shows an X-ray brightness profile in a 120\deg\ sector
centered on the bullet's center of curvature and directed along its apparent
motion. There are two brightness edges whose shapes indicate spherical gas
density discontinuities in projection. We fit this profile by the projection
of a gas density model consisting of two power laws $r^\alpha$ centered on
the bullet representing the bullet gas and the shock region,
respectively. These are immersed in a $\beta$-model centered on the main
cluster (region {\em j}) representing the outer, undisturbed gas. All
components are spherically symmetric around their respective centers; it is
an adequate assumption in the sector of interest. Free parameters are the
three slopes, two jump amplitudes and two jump radii. A reasonable range of
core radii and center positions for the $\beta$-model was explored
(obviously, these cannot be restricted by the fit) and found to have a small
effect on our main interesting parameter, the density jump at the shock. Its
confidence interval includes this modeling uncertainty. The observed
temperature difference has a negligible effect on the derived density.

\noindent
\pspicture(0,8.9)(9,24.3)

\rput[tl]{0}(0.0,24.0){\epsfxsize=8.7cm \epsfclipon
\epsffile{1e_tmap.ps_dist}}

\rput[tl]{0}(-0.04,19.7){\epsfxsize=8.7cm \epsfclipon
\epsffile{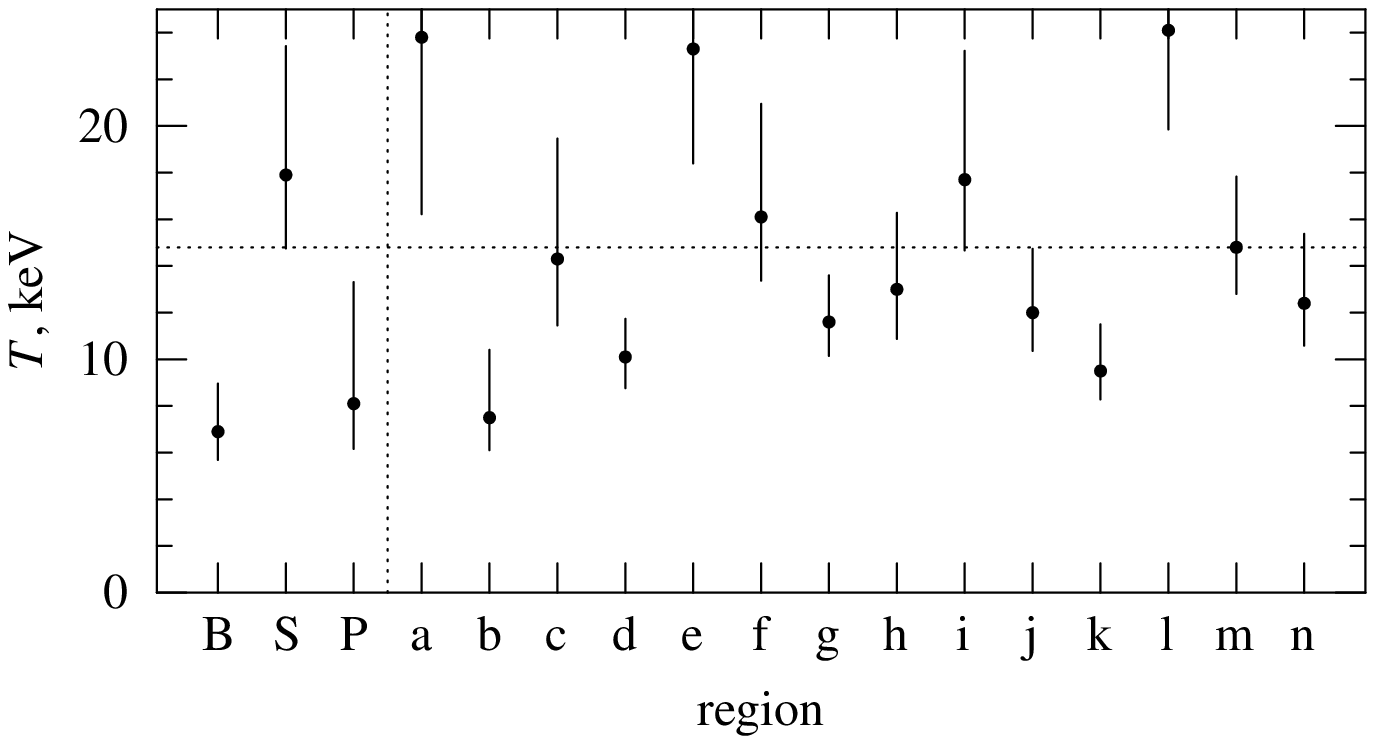}}

\rput[tl]{0}(-0.1,11.3){
\begin{minipage}{8.75cm}
\small\parindent=3.5mm%
{\sc Fig.}~2.---Projected gas temperature map (colors) overlaid on the X-ray
brightness contours from Fig.\ 1{\em b}. Temperatures for individual regions
are shown in lower panel (region {\em B}\/ is the tip of the bullet,
unmarked in the map for clarity.  This region appears smaller than the
bullet in the contour plot because of smoothing). Errors in this figure are
$1\sigma$. The horizontal line shows the cluster average temperature.
\par
\end{minipage}
}
\endpspicture

The best-fit density model is shown in Fig.\ 4{\em b}; it describes the
brightness profile in Fig.\ 4{\em a} well. The best-fit radial slopes are
$\alpha\approx 0.15$ for the bullet and $\alpha\approx -0.3$ for the shock
region; $\beta\approx 0.7$ assuming a core-radius of $125\,h^{-1}$ kpc. The
density jumps by factors of $3.8^{+1.3}_{-1.0}$ at the bullet edge and
$3.2\pm 0.8$ at the shock front. Figure 4{\em b}\/ also shows an approximate
pressure profile (the density model times the temperatures from Fig.\
2). The approximate pressure continuity at the first jump indicates that the
bullet boundary is a ``cold front'', or contact discontinuity, similar to
those recently discovered in other clusters. As expected, there is a large
pressure increase at the shock front.

A density jump of $3.2\pm 0.8$ at the shock corresponds to a Mach number
$M=3.4$ ($>2.1$ at 90\%) for a $\gamma=5/3$ gas and a one-dimensional shock
(strictly speaking, the latter approximation applies only along the
direction of the motion, but we can use it as a qualitative estimate for our
wide sector).  The observed temperature jump from $8^{+9}_{-3}$ keV to
$18^{+9}_{-5}$ keV corresponds to $M=2.1\pm 1.1$ (again, for a qualitative
estimate, we assume constant temperatures in regions {\em P}\/ and {\em
S}). These two independent derivations for the Mach number agree within
their 90\% uncertainties, and we conclude that $M=2-3$. In the stationary
regime, the subcluster should move with the shock velocity. Such Mach
numbers and the observed gas temperatures correspond to $v\sim
3000-4000$\kms, implying that the bullet has passed the center of the main
cluster ($\sim 0.3\,h^{-1}$ Mpc away) just $0.1-0.2$ Gyr ago.

In principle, $M$ could also be estimated from the Mach cone: the asymptotic
angle $\varphi$ of the shock w.r.t.\ the symmetry axis should satisfy
$\sin\varphi = M^{-1}$. For $M=2-3$, $\varphi$ should be $20-30^\circ$,
whereas the image suggests $\varphi\gax 45^\circ$.  A deviation of the
velocity vector from the sky plane could widen the Mach cone in projection;
however, a sufficiently large inclination angle also would smear the
observed sharp brightness edges. An optical measurement of the subcluster
relative line of sight velocity in the comoving frame, $\sim 800$\kms\
(Barrena et al.), combined with our value of $M$, also indicates only a
small ($10-15^\circ$) deviation from the sky plane. The Mach cone relation
assumes a uniform pre-shock medium and constant velocities, and the cone
angle discrepancy is probably due to the cluster's radially declining
density profile and deceleration of the bullet.

\noindent
\pspicture(0,-0.4)(10,8.9)

\rput[bl]{0}(0,0){\epsfxsize=8.7cm \epsfclipon
\epsffile{1e_tmap_radio.ps_dist}}

\rput[bl]{0}(1.9,1.5){\epsfxsize=6.27cm \epsfysize=6.92cm \epsfclipon
\epsffile{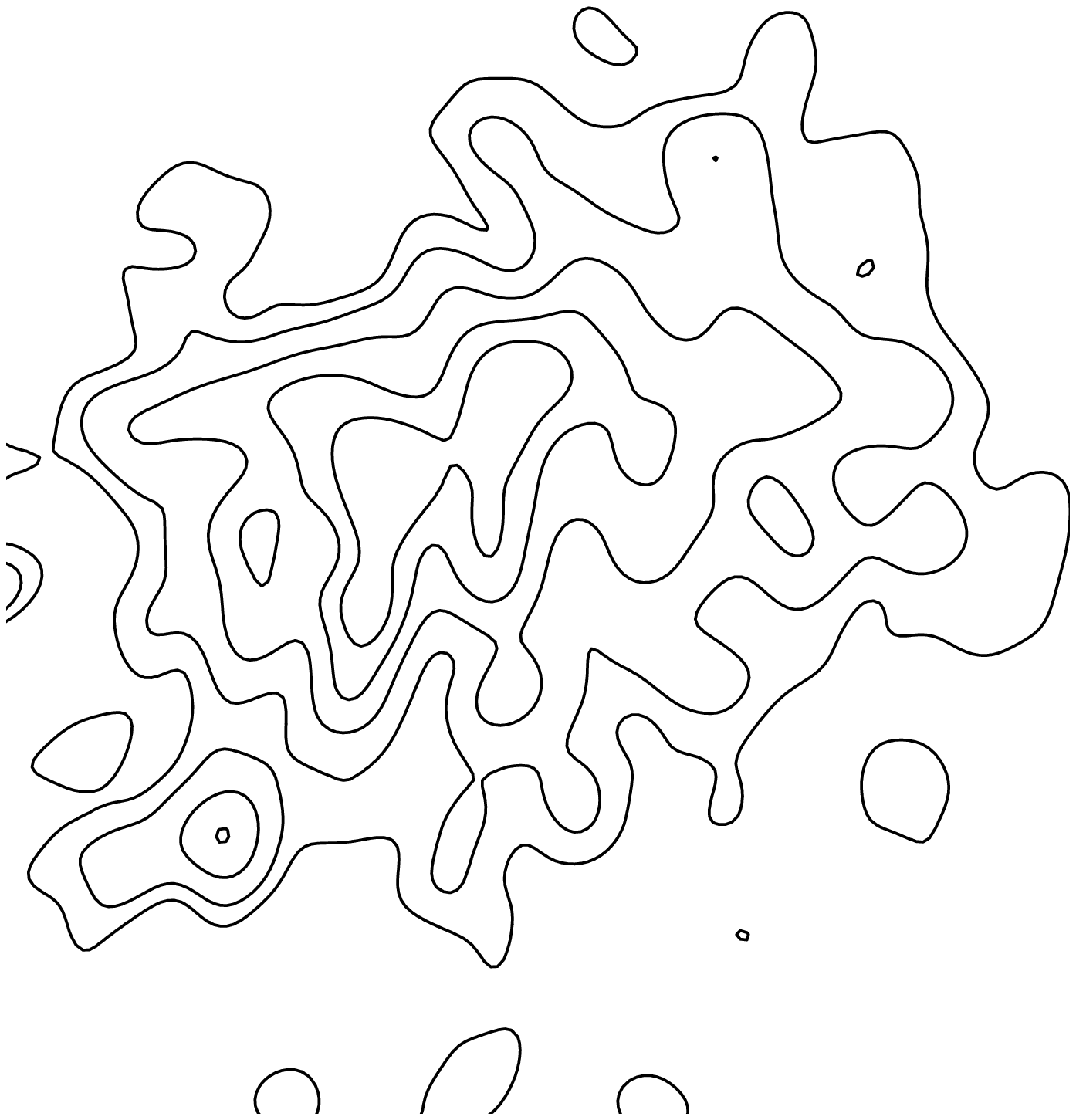}}

\rput[tl]{0}(-0.1,0.5){
\begin{minipage}{8.75cm}
\small\parindent=3.5mm%
{\sc Fig.}~3.---Radio halo brightness contours (reproduced from LHBA)
overlaid on the temperature map from Fig.\ 2. The contours are at $(3, 6,
12, 18, 24) \times \sigma$.
\par
\end{minipage}
}
\endpspicture

\begin{figure*}[t]
\pspicture(0,13.7)(18.5,23.4)

\rput[tl]{0}(0.0,24.0){\epsfysize=9cm \epsfclipon
\epsffile{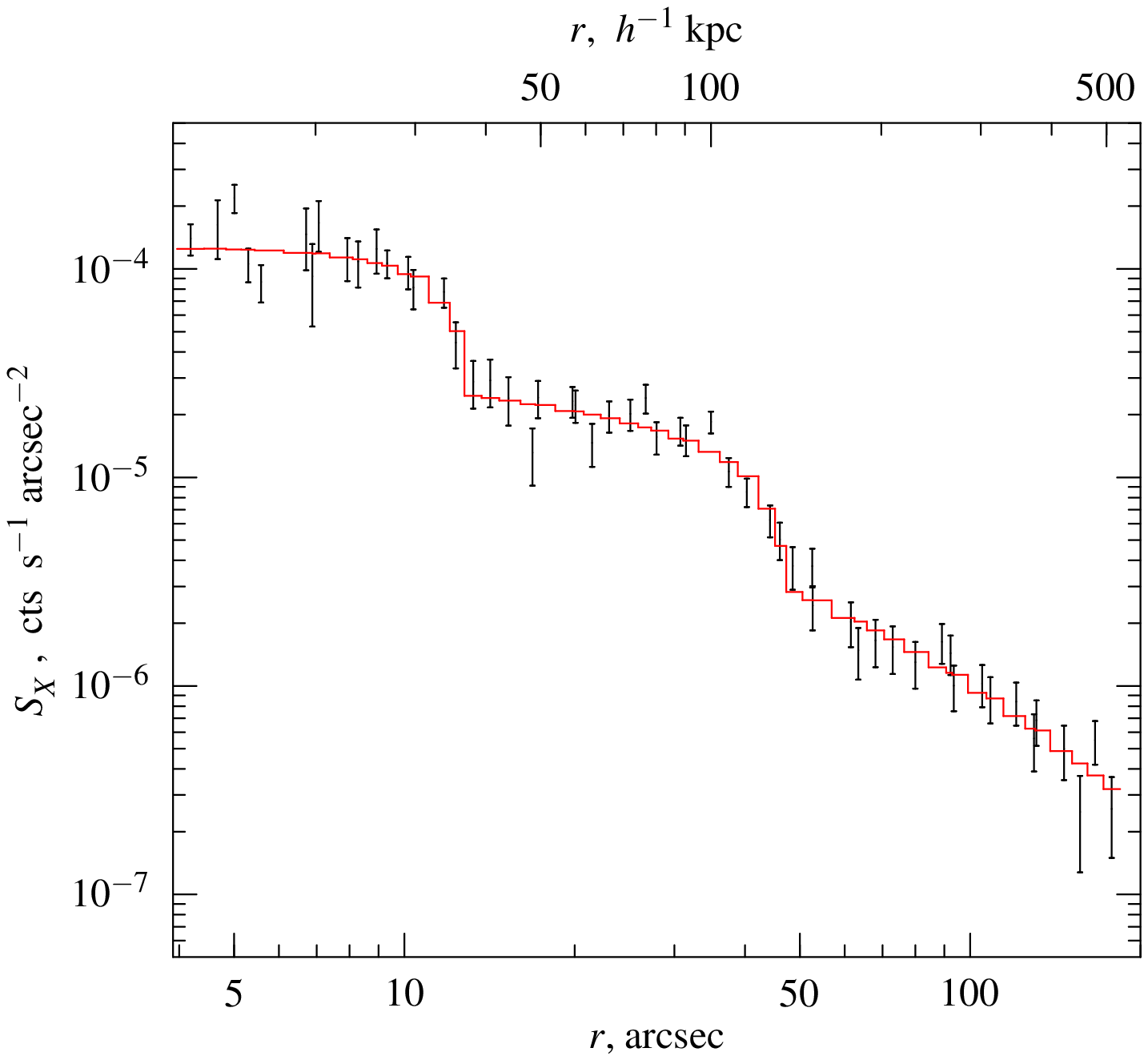}}

\rput[tl]{0}(9.5,24.0){\epsfysize=9cm \epsfclipon
\epsffile{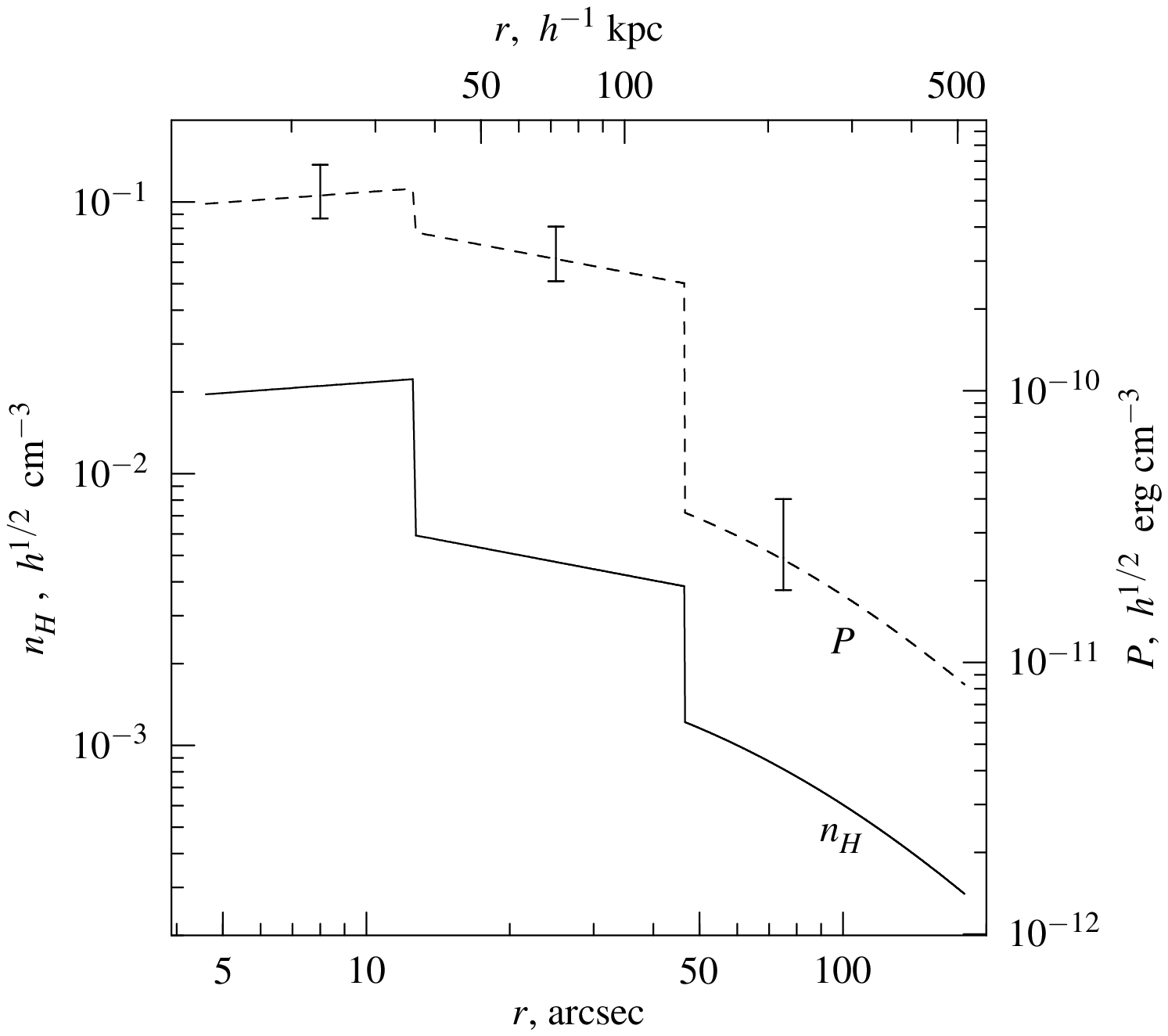}}

\rput[bl]{0}(7.9,21.9){\large\bi a}
\rput[bl]{0}(16.4,21.9){\large\bi b}

\rput[tl]{0}(-0.1,15.3){
\begin{minipage}{18.5cm}
\small\parindent=3.5mm

{\sc Fig.}~4.---({\em a}) ACIS 0.5--5 keV surface brightness profile in a
120\deg\ sector centered on the ``bullet'' and directed westward. The
profile is extracted in elliptical segments parallel to the shock front; the
$r$\/ coordinate corresponds to an average radius within a segment. Errors
in this figure are $1\sigma$. The histogram shows the best-fit model (see
text). The corresponding unprojected model density profile along the
symmetry axis is shown in panel ({\em b}), which also includes an
approximate gas pressure model using temperatures in regions {\em B, S, P}
from Fig.\ 2. Error bars on pressure correspond to errors in temperature.
\par
\end{minipage}
}
\endpspicture
\end{figure*}

\subsection{The bullet subcluster}

The 0.5--5 keV luminosity of the bullet subcluster is $\sim 5\times
10^{43}\, h^{-2}$\ergs, $10-15$\% of the typical value for a cluster with
the bullet's temperature; this fraction would change little if adiabatic
compression of the bullet by the surrounding hot gas is taken into
account. The high gas density in the bullet (Fig.\ 4{\em b}) is of the order
of that in the cooling flow clusters at comparable radii. This suggests that
the bullet is a remnant of a density peak (cooling flow) that once was at
the center of the merging subcluster, perhaps around its brightest galaxy
(Fig.\ 1{\em b}).  The subcluster's less dense outer gas was probably
shocked and ram pressure stripped to the radius where the pressures are
balanced (see Fig.\ 4{\em b}). Most of the stripped gas may reside in the
north-south bar-like structure near the center of the main cluster seen in
Fig.\ 1{\em a}\/ (which may be a pancake in projection), since that is where
the ram pressure on the moving subcluster was the highest. Disruption of a
cooling flow by a merger was considered theoretically by, e.g., Fabian \&
Daines (1991) and G\'omez et al.\ (2000).

The shuttlecock shape of the bullet clearly shows that it continues to
be actively destroyed by gas dynamic instabilities (e.g., Jones, Ryu, \&
Tregillis 1996). The gas being swept back from the cool bullet appears to be
quite hot (region {\em a} in Fig.\ 2) and clumpy (regions {\em a, b}),
suggesting interesting physics at the interface of the two gases. This
interface will be studied in detail using a longer \chandra\ observation
planned for 2002.

\subsection{High overall temperature}

The existence of a few extremely hot clusters such as \1e\ have been used in
the past to derive cosmological constraints under the assumption that the
high temperature indicates high virial mass. However, the results presented
here show that the assumption of hydrostatic equilibrium in \1e\ can easily
overestimate the mass by a significant factor due to the ongoing merger and
a temporary increase in temperature (see, e.g., simulations by Ricker \&
Sarazin 2001 and Ritchie \& Thomas 2001). It is thus difficult to draw any
strong cosmological conclusions without a better estimate of the cluster
mass.

\subsection{Possible future measurements}

The unique shock and supersonic subcluster in \1e\ enable several
interesting measurements. Since the gas density and temperature jumps at the
shock are related, their accurate measurement may give the equation of state
for the intracluster plasma. Magnetic fields, a relativistic particle
population with sufficient pressure, a significant lag between the electron
and ion temperatures, and large nonthermal energy losses in a shock could
all change the true or the apparent value of $\gamma$.

From the velocity of the bullet and the density of the ambient gas, we can
derive the ram pressure on the bullet. The gas bullet appears to be pushed
away from the subcluster's dark matter potential well just now, so the
gravitational pull of the subcluster should equal the ram pressure. This may
give an independent estimate of the subcluster's total mass (Markevitch et
al.\ 1999). 

The cluster also presents an interesting opportunity to constrain the
collisional nature of dark matter (e.g., Spergel \& Steinhardt 2000;
Furlanetto \& Loeb 2001). There is a clear offset between the centroid of
the bullet subcluster's galaxies and its gas. If one measures the location
of the subcluster's dark matter density peak (e.g., from weak lensing or
detailed modeling of the gas distribution), one may determine whether the
dark matter is collisionless, as are the galaxies, or if it experiences an
analog of ram pressure, as does the gas.

\section{SUMMARY}

The \chandra\ observation of \1e\ presents a prototypical example of a
merger bow shock. The shock propagates in front of a cooler gas ``bullet'',
apparently a remnant of a dense central region of the merging subcluster
whose outer gas was stripped by ram pressure. The subcluster Mach number is
2--3 and its velocity is 3000--4000\kms. Thus the subcluster traversed the
main cluster core only 0.1--0.2 Gyr ago. The bullet is at the final stage of
being destroyed by gas dynamic instabilities. Its gas lags behind the
subcluster galaxies due to the ram pressure of the shocked gas. The hottest
gas resides in a different region of \1e\ where additional merging activity
occurs. The overall high temperature of \1e\ is unlikely to represent its
virial temperature due to the ongoing merger.

\acknowledgements

We thank L. Grego, H. Tananbaum and the referee for useful comments. Support
was provided by NASA contract NAS8-39073, grant NAG5-9217, the Smithsonian
and the CfA fellowship program.


\begin{references}




\reference{} Barrena, R., Biviano, A., Ramella, M., Falco, E., \& Seitz,
S. 2001, Tracing Cosmic Evolution with Galaxy Clusters, eds.\ S. Borgani et
al., ASP Conference Series, in press

\reference{} Dickey, J. M., \& Lockman, F. J. 1990, ARA\&A, 28, 215


\reference{} Fabian, A.C., \& Daines, S. J. 1991, MNRAS, 252, 17P

\reference{} Furlanetto, S. R., \& Loeb, A. 2001, ApJ, submitted
(astro-ph/0107567)

\reference{} Furusho, T., Yamasaki, N., Ohashi, T., Shibata, R., \& Ezawa,
H. 2001, ApJL, in press (astro-ph/0110146)

\reference{} G\'omez, P.L., Loken, C., Roettiger, K., \& Burns, J. O. 2000,
ApJ, in press (astro-ph/0009465)

\reference{} Govoni, F., En{\ss}lin, T. A., Feretti, L., \& Giovannini, G.
2001, A\&A 369, 441

\reference{} Jones, T.~W., Ryu, D., \& Tregillis, I.~L. 1996, ApJ, 473, 365

\reference{} Liang, H., Hunstead, R. W., Birkinshaw, M., \& Andreani, P.
2000, ApJ, 544, 686 (LHBA)

\reference{} Henry, J. P., \& Briel, U. G. 1995, ApJ, 443, L9


\reference{} Kaastra, J. S. 1992, ``An X-Ray Spectral Code for Optically Thin
Plasmas'' (Internal SRON-Leiden Report, updated version 2.0)

\reference{} Markevitch, M., et al.\ 2000, ApJ, 541, 542 

\reference{} Markevitch, M., Sarazin, C. L., \& Vikhlinin, A. 1999, ApJ,
521, 526

\reference{} Markevitch, M., \& Vikhlinin, A. 2001, ApJ, in press;
astro-ph/0105093 (MV)

\reference{} Neumann, D. M., et al.\ 2001, A\&A, 365, L74

\reference{} Ricker, P. M., \& Sarazin, C. L. 2001, ApJ in press
(astro-ph/0107210)

\reference{} Ritchie, B. W., \& Thomas, P. A. 2001, MNRAS, submitted
(astro-ph/0107374)


\reference{} Sarazin, C. L. 2001, Galaxy Clusters and the High Redshift
Universe Observed in X-rays, XXXVI Recontres de Moriond, in press
(astro-ph/0105458)

\reference{} Shafranov, V. D. 1957, Soviet Phys.\ JETP, 5, 1183

\reference{} Spergel, D. N., \& Steinhardt, P. J.\ 2000, Physical Review
Letters, 84, 3760

\reference{} Takizawa, M. 1999, ApJ, 520, 514 


\reference{} Tribble, P. 1993, MNRAS, 263, 31

\reference{} Tucker, W.~H., Tananbaum, H., \& Remillard, R.~A.\ 1995, ApJ,
444, 532

\reference{} Tucker, W., et al.\ 1998, ApJ, 496, L5 (T98)

\reference{} Vikhlinin, A., Markevitch, M., \& Murray, S. S. 2001, ApJ, 551,
160

\reference{} Yaqoob, T.\ 1999, ApJ, 511, L75 


\end{references}
\end{document}